\documentclass[12pt]{article}
\usepackage{amsfonts}
\usepackage{epsfig}

\begin{document}

\title{Dilepton decays of nucleon resonances\\
and dilepton production cross sections \\
in proton-proton collisions}
\author{M.I. Krivoruchenko$^{1,2)}$ \\
%EndAName
$^{1)}${\small Institut f\"{u}r Theoretische Physik, Universit\"{a}t
Tuebingen, Auf der Morgenstelle 14 }\\
{\small D-72076 T\"{u}bingen, Germany }\\
{\small \ }$^{2)}${\small Institute for Theoretical and Experimental
Physics, B. Cheremushkinskaya 25 }\\
{\small 117259 Moscow, Russia}}
\maketitle

\begin{abstract}
Relativistic, kinematically complete phenomenological expressions for the
dilepton decay rates of nucleon resonances with arbitrary spin and parity
are derived in terms of the magnetic, electric, and Coulomb transition form
factors. The dilepton decay rates of the nucleon resonances with masses
below $2$ GeV are estimated using the extended vector meson dominance (eVMD)
model for the transition form factors. The model provides an unified
description of the photo- and electroproduction data and of the vector meson
and dilepton decays of the nucleon resonances. The constraints on the
transition form factors from the quark counting rules are taken into account
explicitly. The remaining parameters of the model are fixed by fitting the
available photo- and electroproduction data and using results of the
multichannel partial-wave analysis of the $\pi N$ scattering. The results
are used to describe dilepton spectra measured at BEVALAC in proton-proton
collisions.
\end{abstract}

\section{Introduction}
Dileptons are the clearest probe to study highly compressed nuclear matter. 
They provide a possibility to measure 
experimentally the in-medium widths and masses of vector mesons. 
The dilepton spectra measured by the CERES \cite{ceres} 
and HELIOS-3 \cite{HELIOS} Collaborations at CERN SPS found
a significant enhancement of the low-energy dilepton yield below the 
$\rho $ and $\omega $ peaks in heavy systems 
($Pb+Au$) compared to light systems ($S+W$) and 
proton induced reactions ($p+Be$). Theoretically, this enhancement 
can be explained in a hadronic picture assuming a dropping mass 
scenario for the $\rho $ meson or by the inclusion 
of in-medium spectral functions for the vector mesons. In both cases 
the enhanced low energetic 
dilepton yield is not simply due to a shift of the 
$\rho $ and $\omega $ peaks in the nuclear medium but it originates 
to most extent from an enhanced contribution of the $\pi^+ \pi^{\;-}$ annihilation 
channel which, assuming vector dominance, runs over an intermediate 
in-medium $\rho$ mesons. An alternative scenario is the formation 
of a quark-gluon plasma in the heavy systems which leads to additional 
contributions to the dilepton spectrum from perturbative QCD ($pQCD$) 
such as quark-antiquark annihilation \cite{rapp,weise00}.

Concerning the DLS experiment \cite{BEVALAC}, the measured dilepton spectra do not 
match with the theoretical estimates, even when possible
reduction of the $\rho $-meson mass and the $\rho $-meson broadening are
taken into account \cite{Bratkovskaya:1999pr}. This phenomenon is called 'DLS puzzle'. 
The HADES experiment at GSI will study the 
dilepton spectra in the same energy range in greater details \cite{Friese:1999qm}. 

The experimental data from heavy-ion collisions can only be compared to
theoretical predictions from transport models which account for the 
complicated reaction dynamics. 
The elementary cross sections enter as an input into the transport simulations 
of heavy-ion collisions. A precise and rather complete knowledge of the 
decay channels of mesons and nucleon resonances, and
their production, absorption, and reabsorption cross sections is therefore 
indispensable in order to draw reliable conclusions from studies of the dilepton 
production. The recent measurement of the dilepton production in proton-proton 
collisions by BEVALAC \cite{W} provides a useful tool for testing the current
theoretical schemes.

We give here relativistic, kinematically complete phenomenological expressions 
for the dilepton decay rates of nucleon resonances with arbitrary spin and parity
and discuss these results in connection to the dilepton production in 
$pp$ collisions at BEVALAC energies. 

\section{Dilepton widths}

The signs $\pm$ stand
for the normal- and abnormal parity resonances, $J^{P}=\frac{1}{2}^{-},\frac{%
3}{2}^{+},\frac{5}{2}^{-},\;...$ (the upper sign) and $J^{P}=\frac{1}{2}^{+},%
\frac{3}{2}^{-},\frac{5}{2}^{+},...$ (the lower sign). In terms of the magnetic ($M$), 
electric ($E$), and Coulomb ($C$) form factors, the
decay width of a nucleon resonance with spin $J=l+\frac{1}{2} \ge \frac{3}{2}$ and mass $m_*$ 
into the nucleon with mass $m$ and a virtual photon with mass $M$
equals \cite{krivo02}
\begin{eqnarray}
\Gamma (N_{(\pm )}^{*} &\rightarrow &N\gamma ^{*})=\frac{9\alpha }{16}\frac{%
(l!)^{2}}{2^{l}(2l+1)!}\frac{m_{\pm }^{2}(m_{\mp }^{2}-M^{2})^{l+1/2}(m_{\pm
}^{2}-M^{2})^{l-1/2}}{m_{*}^{2l+1}m^{2}}  \nonumber \\
&&\left( \frac{l+1}{l}\left| G_{M/E}^{(\pm )}\right| ^{2}+(l+1)(l+2)\left|
G_{E/M}^{(\pm )}\right| ^{2}+\frac{M^{2}}{m_{*}^{2}}\left| G_{C}^{(\pm
)}\right| ^{2}\right)  \label{GAMMA_l}
\end{eqnarray}
where $m_{\pm}=m_* \pm m$ and $G^{\pm}_{E/M}$ means $G^{+}_{E}$ or $G^{-}_{M}$. 
For $l=1$, we recover the result of ref. \cite{Krivoruchenko:2001hs}.

The resonance decay widths of $J=\frac{1}{2}$ nucleon resonances can be found to be 
\begin{eqnarray}
\Gamma (N_{(\pm )}^{*} &\rightarrow &N\gamma ^{*})=\frac{\alpha }{8m_{*}}%
(m_{\pm }^{2}-M^{2})^{3/2}(m_{\mp }^{2}-M^{2})^{1/2}  \nonumber \\
&&\left( 2\left| G_{E/M}^{(\pm )}\right| ^{2}+\frac{M^{2}}{m_{*}^{2}}\left|
G_{C}^{(\pm )}\right| ^{2}\right) .  \label{GAMMA_0}
\end{eqnarray}

We use here the normalization for the monopole form factors identical to refs. \cite
{krivo02,Devenish:1976jd}. The $\Delta (1232)$-resonance form factors of refs. \cite
{Krivoruchenko:2001hs,JS} contain an additional factor of $\sqrt{\frac{2}{3}}%
.$ 

If the width $\Gamma (N^{*}\rightarrow N\gamma ^{*})$ is known, the
factorization prescription (see {\it e.g.} \cite{Faessler:2000de}) can be
used to find the dilepton decay rate:

\begin{equation}
d\Gamma (N^{*}\rightarrow Ne^{+}e^{-})=\Gamma (N^{*}\rightarrow N\gamma
^{*})M\Gamma (\gamma ^{*}\rightarrow e^{+}e^{-})\frac{dM^{2}}{\pi M^{4}},
\label{OK!}
\end{equation}
where 
\begin{equation}
M\Gamma (\gamma ^{*}\rightarrow e^{+}e^{-})=\frac{\alpha }{3}%
(M^{2}+2m_{e}^{2})\sqrt{1-\frac{4m_{e}^{2}}{M^{2}}}  \label{OK!!}
\end{equation}
is the decay width of a virtual photon $\gamma ^{*}$ into the dilepton
pair with invariant mass $M$. Eqs.(1)-(\ref{OK!!}) being combined give 
the $N^{*}\rightarrow Ne^{+}e^{-}$ decay rates. 

The $\Delta (1232)$ dilepton decays are known as the dominant sources of
the dilepton yield in nucleon-nucleon and heavy-ion collisions at low energies. 
As we discussed in ref. \cite{Krivoruchenko:2001hs}, previous calculations of the 
$\Delta(1232)\rightarrow Ne^{+}e^{-}$ decays, available in the literature, 
are incorrect. 

\section{Extended VMD $\&$ quark counting rules}

To proceed further, one needs a specific model for transition form factors 
of nucleon resonances. We use extended VMD (eVMD) model. It is motivated by 
two observations:

(i) The naive VMD model should give, in principle, an unified description 
of the radiative $R \rightarrow N\gamma $ and the mesonic $R \rightarrow NV$ decays. 
However, a normalization to the radiative branchings ($RN\gamma $) strongly 
underestimates the mesonic branchings ($RNV$)
as discussed in refs. \cite{friman97,Faessler:2000md}.

(ii) The electromagnetic nucleon form factors demonstrate experimentally a dipole
behavior. The quark counting rules for the Sachs form factors predict $%
G_{E}(q^{2})\sim G_{M}(q^{2})\sim 1/q^{4}$ at $q^{2}\rightarrow
\infty .$ The naive VMD model with the ground-state $\rho $-$,$ $\omega $-$,$ and $%
\phi $-mesons cannot describe quantitatively the nucleon form factors and gives 
incorrect asymptotic behavior. It was proposed \cite
{Hohler:1976ax} to
include in the electromagnetic current excited states of the vector mesons $\rho
^{\prime },$ $\rho ^{\prime \prime },$ ... etc. 

The eVMD allows to solve naturally the problem of the $RN\gamma$ to $RNV$ ratios.
The requirement of a stronger suppression of the transition form factors at high $q^2$ is
equivalent to a {\it destructive interference} of the $\rho$- and $\omega$-families
away from the $\rho$ and $\omega$ poles. It reduces the $RN\gamma$ to $RNV$ ratios.

The monopole transition form factors, $G^{\pm}_T(M^2)$ with $T=M,E,C$, are expressed in terms of the 
covariant form factors, $F^{\pm}_k(M^2)$ with $k=1,2,3$, as follows

\begin{equation}
G_{T}^{(\pm)}(M^{2})=\sum_{k}{\rm M}_{Tk}(M^{2})F_{k}^{(\pm)}(M^{2}).  \label{DRUN}
\end{equation}
The transformation matrices ${\rm M}_{Tk}$ can be found in ref. \cite{JS} for $J^P= \frac{3}{2}^+$ 
and in refs. \cite{krivo02,Devenish:1976jd} for arbitrary $J^P$. 

The quark counting rules \cite{Matveev:1973ra} predict the following asymptotics for the
covariant form factors of $J\geq \frac{3}{2}$ nucleon resonances:

\begin{eqnarray}
F_{1}^{(\pm )}(M^{2}) &=&O(\frac{1}{(-M^{2})^{l+2}}), \nonumber \\
F_{2}^{(\pm )}(M^{2}) &=&O(\frac{1}{(-M^{2})^{l+3}}), \nonumber \\
F_{3}^{(\pm )}(M^{2}) &=&O(\frac{1}{(-M^{2})^{l+3}}).
\end{eqnarray}

In the no-widths approximation for the vector mesons, these constraints can be 
resolved to give
\begin{eqnarray}
F_{1}^{(\pm )}(M^{2}) &=&\frac{{\sum }_{j=0}^{n+1}C_{1j}^{(\pm )}{M^{2}}^{j}%
}{{\prod }_{i=1}^{l+3+n}(1-M^{2}/m_{i}^{2})},  \nonumber \\
F_{2}^{(\pm )}(M^{2}) &=&\frac{{\sum }_{j=0}^{n}C_{2j}^{(\pm )}{M^{2}}^{j}}{{%
\prod }_{i=1}^{l+3+n}(1-M^{2}/m_{i}^{2})},  \nonumber \\
F_{3}^{(\pm )}(M^{2}) &=&\frac{{\sum }_{j=0}^{n}C_{3j}^{(\pm )}{M^{2}}^{j}}{{%
\prod }_{i=1}^{l+3+n}(1-M^{2}/m_{i}^{2})}.  \label{FF_l}
\end{eqnarray}
Here, $C_{kj}^{(\pm )}$ are free parameters of the eVMD, ${%
l+3+n}$ is the total number of the vector mesons with masses $m_i$. The
quark counting rules reduce the number of free parameters from ${l+3+n}$ to $%
{n+2}$ for $k=1$ and to ${n+1}${\ for} $k=2,3$. In the minimal case $n=0,$
the knowledge of the four parameters $C_{10}^{(\pm )}$, $C_{11}^{(\pm )}$, 
$C_{20}^{(\pm )}$, and $C_{30}^{(\pm )}$ is sufficient to fix $F_{k}^{(\pm
)}(M^{2})$. In the zero-width limit, the multiplicative representation (\ref
{FF_l}) is completely equivalent to the usual additive representation.

The similar multiplicative representation motivated by the Regge theory
is used in ref. \cite{Devenish:1976jd}. The asymptotic dominance of the transverse 
covariant form factors, used in that work as an assumption, however,
does not agree with the quark counting rules. 

For spin $J=\frac{1}{2}$ resonances, the constraints have the form
\begin{equation}
F_{1,2}^{(\pm )}(M^{2})=O(\frac{1}{(-M^{2})^{3}}).
\end{equation}
The general representation for the covariant form factors in the spin-$\frac{1}{2}$
case becomes
\begin{equation}
F_{k}^{(\pm )}(M^{2})=\frac{{\sum }_{j=0}^{n}C_{kj}^{(\pm )}{M^{2}}^{j}}{{%
\prod }_{i=1}^{3+n}(1-M^{2}/m_{i}^{2})}.  \label{FF_0}
\end{equation}

The free parameters of the eVMD model are fixed by fitting the available
photo- and electroproduction data, $\gamma(\gamma ^{*})N
\rightarrow N^{*},$ and the vector meson decays, $N^{*}\rightarrow
N\rho(\omega) $. When the experimental data are not available, the
quark model predictions are used as an input.

%%%%%%%%%%%%%%%%%%%%%%%%%%%%%%%%%%%%%%%%%%%%%%%%%%%%%%%%%%%%%%%%%%%
\section{Dilepton production in proton-proton collisions}
%%%%%%%%%%%%%%%%%%%%%%%%%%%%%%%%%%%%%%%%%%%%%%%%%%%%%%%%%%%%%%%%%%%
A possibility to clarify the origin of the DLS puzzle has appeared since
data from elementary $pp$ $(pd)$ collisions at $%
T=1\div 5$ GeV ($T$ is the kinetic energy of the incident proton in the
laboratory frame) became available from the DLS Collaboration \cite{W}. 

For description of the dilepton production in proton-proton collisions at
energies $1 \div 3$ GeV, we use the nucleon resonance model.
The mesons $P$ $(= \pi, \eta, ...)$ 
and $V$ $(= \rho, \omega, \phi)$ are produced through 
a two-step mechanism via the excitation of nuclear resonances, i.e. 
$NN \rightarrow NR$, $R \rightarrow NV$. When energy increases, 
multiparticle finals states become dominant. For such energies we used
the experimental inclusive cross sections for the meson production. 

The $pp\rightarrow ppM$ cross section with $M=P,V$ is given by
\begin{equation}
\frac{d\sigma (s,M)^{pp\rightarrow ppM}}{dM^{2}}
=\sum_{R}\int_{(m_{p}+M)^{2}}^{(\sqrt{s}-m_{p})^{2}}d\mu ^{2}
\frac{ d\sigma (s,\mu )^{pp\rightarrow pR}}{d\mu ^{2}}
\frac{dB(\mu,M)^{R\rightarrow pM }}{d M^{2}}.
\label{sigNNV}
\end{equation}
The cross sections for the resonance production are given by 
\begin{equation}
%d\sigma (s,\mu )^{pp\rightarrow pR} = 
%\frac{|{\cal M}_R|^2 ~ p^*(\sqrt {s},\mu,m_{p})}{16 p_i s\pi^2} 
%\frac{\mu \Gamma_{\rm tot}^R (\mu) d\mu^2 }
%{(\mu^2-m_{R}^2)^2 +(\mu\Gamma_{\rm tot}^R(\mu))^2},
d\sigma (s,\mu )^{pp\rightarrow pR} = 
\frac{|{\cal M}_R|^2}{16 p_i \sqrt s \pi^2}\Phi_2(\sqrt {s},\mu,m_{p})dW_R(\mu)
\label{sigNR}
\end{equation}
with $\Phi_2(\sqrt {s},\mu,m_{p}) = \pi p^*(\sqrt {s},\mu,m_{p})/{\sqrt s}$ 
being the two-body phase space, 
$p^*(\sqrt {s},\mu,m_{p})$ the final c.m. momentum, $p_i$ the 
initial c.m. momentum, and $\mu$ and $m_R$ the running and pole masses 
of the resonances, respectively, and $m_p$ is the proton mass. The
mass distribution $dW_{R}(\mu)$ of the resonances is described by the 
standard Breit-Wigner formula. The sum in (\ref{sigNNV}) runs over the well 
established (4$*$) resonances quoted by the PDG \cite{PDG}. 
The branching to the $V$ decay mode is given by 
\begin{equation}
dB(\mu,M)^{R\rightarrow pV } = 
\frac{d\Gamma_{\rm NV}^R (\mu,M)}{\Gamma_R (\mu)}.
\label{bra}
\end{equation}

In terms of
the magnetic, electric, and Coulomb couplings $g_M^{(\pm)}$, $g_E^{(\pm)}$, 
and $g_C^{(\pm)}$, the differential
decay widths of nucleon resonances with spin 
$J = l + 1/2$ into a vector meson, $V$, with arbitrary mass $M$ 
has the form \cite{krivo02}
\begin{eqnarray}
d\Gamma_{NV}^R(\mu,M) =\frac{9}{64\pi}\frac{%
(l!)^{2}}{2^{l}(2l+1)!}\frac{m_{\pm }^{2}(m_{\mp }^{2}-M^{2})^{l+1/2}(m_{\pm
}^{2}-M^{2})^{l-1/2}}{\mu^{2l+1}m^{2}}  \nonumber \\
\left( \frac{l+1}{l}\left| g_{M/E}^{(\pm )}\right| ^{2}+(l+1)(l+2)\left|
g_{E/M}^{(\pm )}\right| ^{2}+\frac{M^{2}}{\mu^{2}}\left| g_{C}^{(\pm
)}\right| ^{2}\right)dW_{V}(M),  \label{GAMMA_ll}
\end{eqnarray}
with $m_{\pm} = \mu \pm m_p$. The signs $\pm$ refer to the natural parity and abnormal parity, 
$g_{M/E}^{\pm}$ means $g_{M}^{+}$ or $g_{E}^{-}$. 
The above equation is valid
for $J \geq \frac{3}{2}$. For $J = \frac{1}{2}$ one obtains 
\begin{eqnarray}
d\Gamma_{NV}^R(\mu,M) =\frac{1 }{32\pi\mu}%
(m_{\pm }^{2}-M^{2})^{3/2}(m_{\mp }^{2}-M^{2})^{1/2}  \nonumber \\
\left( 2\left| g_{E/M}^{(\pm )}\right| ^{2}+\frac{M^{2}}{\mu^{2}}\left|
g_{C}^{(\pm )}\right| ^{2}\right)dW_{V}(M).  \label{GAMMA_01}
\end{eqnarray}
%%%%%%%%%%%%%%%%%%%%%%%%%%%%
\begin{figure}
\begin{center}
\leavevmode
\epsfxsize = 17cm
\epsffile[80 40 580 600]{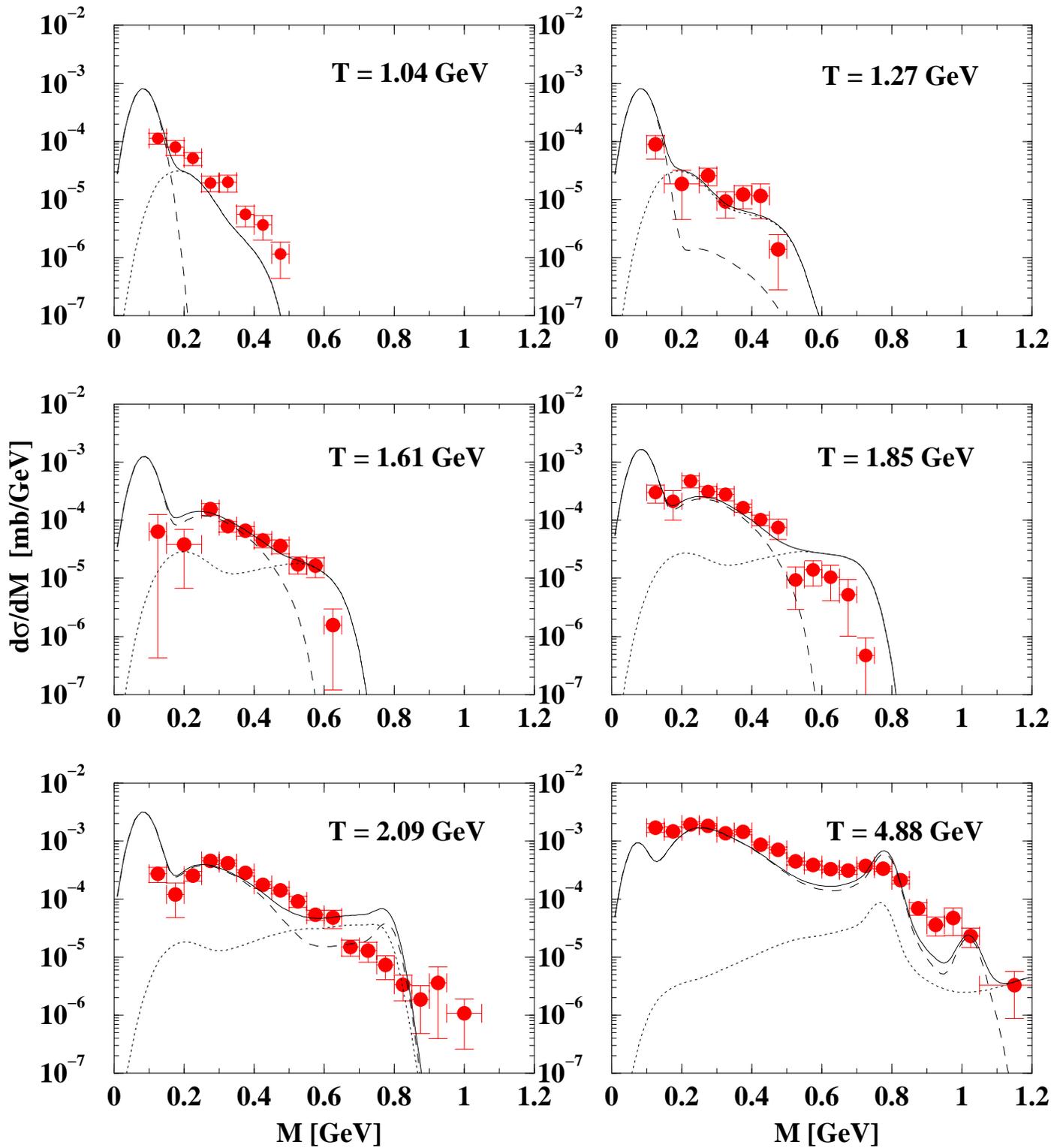}
\end{center}
\caption{The differential dilepton production cross sections as a
function of the dilepton invariant mass, $M$, after applying the
experimental filter and the smearing procedure. The solid
curves are the total cross sections, the dashed curves correspond to the
inclusive production, and the dotted curves correspond to the subthreshold
production. The experimental data are from ref. [8].
}
\label{fig3}
\end{figure}
%%%%%%%%%%%%%%%%%%%%%%%%%%%%%
The distribution $dW_{V}(M)$ is also the Breit-Wigner distribution. 
The last two equations are similar to eqs.(1) and (2) for the virtual photon decays.

Due to the subthreshold character of the $\omega$ production in decays of 
on-shell nucleon resonances, the $M$-dependence of the coupling constants 
$g_M^{(\pm)}$, $g_E^{(\pm)}$, and $g_C^{(\pm)}$ can be important. At 
the $\omega$ pole mass $m_{\omega}$ these couplings are proportional to residues of 
the magnetic, electric, and Coulomb transition form factors. 
We assume that the coupling constants which enter into the covariant 
representation of the form factors are not mass dependent. 
The $M$-dependence of $g_M^{(\pm)}$, $g_E^{(\pm)}$, 
and $g_C^{(\pm)}$ arises then exclusively from the $M$-dependent 
transformation from the covariant basis to the multipole basis according to
%%%%%%%%%%%%%%%%%%5
\begin{equation}
g_{T}^{(\pm)}(M^{2})=\sum_{kT^{\prime }}{\rm M}_{Tk}(M^{2}){\rm M}_{kT^{\prime
}}^{-1}(m_{\omega}^{2})g_{T^{\prime }}^{(\pm)}(m_{\omega}^{2}),  \label{RUN}
\end{equation}
with $T,T^{\prime}=M,E,C$. The dilepton spectra are shown in Fig. 1.
More details on the calculations can be found in ref. \cite{Faessler:2000md}.

\section{Conclusions}

We have considered the dilepton production in $pp$ collisions at BEVALAC
energies $T=1\div 5$ GeV. The subthreshold production of vector mesons through the nucleon
resonances is described within the eVMD model which allows to bring
the transition form factors in agreement with the quark counting rules and
provides an unified
description of the photo- and electroproduction data, $\gamma(\gamma ^{*})N
\rightarrow N^{*},$ the vector meson decays, $N^{*}\rightarrow
N\rho(\omega) $, and the dilepton decays, $N^{*}\rightarrow N\ell ^{+}\ell
^{-}$. The dilepton decay rates are described relativistically using 
kinematically complete phenomenological expressions.

The resulting dilepton spectra are reasonably well described at proton energies 
of $T=1.27\div 1.85$GeV. At $T=1.04$, $T=2.09$, and $T=4.88~$GeV 
the agreement is not perfect. The future experimental investigations at GSI 
will probably shed new light on these problems.

\section{Acknowledgments}
The present contribution is based on works made in collaboration with 
Amand Faessler, C. Fuchs, and B.V. Martemyanov. 
The author is grateful to the Organizing Commettee for financial support.

\end{document}